\documentstyle[epsfig, rotating]{jaa}

\DeclareMathAlphabet{\mathsc}{OT1}{cmr}{m}{sc}
\def\testbx{bx}%
\DeclareRobustCommand{\ion}[2]{%
\relax\ifmmode
\ifx\testbx\f@series
{\mathbf{#1\,\mathsc{#2}}}\else
{\mathrm{#1\,\mathsc{#2}}}\fi
\else\textup{#1\,{\mdseries\textsc{#2}}}%
\fi}
\def\ch{\footnotesize}
\def\HI{\ion{H}{i}~}
\def\km{km~s$^{-1}$~}
\def\deg{\hbox{$^\circ$}~}

\def\AJ{Astron. J.}

\begin{document}
\title[Eridanus group : Tully-Fisher relations]
{The Tully-Fisher relations of the Eridanus group of galaxies}
\author[Omar \& Dwarakanath] 
{A. Omar$^{1}$\thanks{e-mail: aomar@upso.ernet.in}
\& K.S. Dwarakanath$^{2}$\thanks{e-mail: dwaraka@rri.res.in}\\
(1) Aryabhatta Research Institute of observational-sciencES, Manora peak, Nainital 263 129, India\\
(2) Raman Research Institute, Sadashivanagar, Bangalore 560 080, India 
}

\pubyear{xxxx}
\volume{xx}
\date{Received xxx; accepted xxx}
\maketitle
\label{firstpage}
\begin{abstract}

The Tully-Fisher (TF) or the luminosity line-width relations of the galaxies in the
Eridanus group are constructed using the \HI rotation curves and the luminosities in
the optical and in the near-infrared bands. The slopes of the TF relations (absolute
magnitude $vs$ log$2$V$_{flat}$) are $-8.6\pm1.1$, $-10.0\pm1.5$, $-10.7\pm2.1$, and
$-9.7\pm1.3$ in the R, J, H, and K bands respectively for galaxies having flat \HI
rotation curves. These values of the slopes are consistent with those obtained from
studies of other groups and clusters. The scatter in the TF relations is in the range
$0.5 - 1.1$~mag in different bands. This scatter is considerably larger compared to
those observed in other groups and clusters. It is suggested that the larger scatter
in the TF relations for the Eridanus group is related to the loose structure of the
group. If the TF relations are constructed using the baryonic mass (stellar + \HI +
Helium mass) instead of the stellar luminosity, nearly identical slopes are obtained
in the R and in the near-infrared bands. The baryonic TF (baryonic mass $vs$
log$2$V$_{flat}$) slope is in the range $3.5 - 4.1$.  \end{abstract} 


\section{Introduction}

A correlation between the total mass derived using the HI line-widths and optical
luminosity was noticed in the late 70's and early 80's by a number of investigators
(e.g., Roberts 1969, Rogstad \& Shostak 1972, Balkowski et al. 1974, Shostak 1975).
Tully \& Fisher (1977) suggested a luminosity-linewidth relation of the form $L
\propto V_{rot}^{\alpha}$ to measure distances to spiral galaxies. Subsequently, the
luminosity-linewidth relation was also known as the Tully-Fisher (TF) relation. Tully
\& Fisher (1977) demonstrated that using some nearby spiral galaxies for which the
distances are known accurately, the luminosity-linewidth relation can be calibrated to
infer distances to clusters of galaxies as well as to obtain a value for the Hubble
constant. The value of index $\alpha$ has been found to be $\sim 3$ in the optical
bands and $\sim 4$ in the near-infrared bands.

The Tully-Fisher relation can possibly be best attributed to a tight correlation
between dark matter and the total baryonic matter embedded in it. This is because the
constant rotation velocity in the outer regions of galaxies is believed to be due to
dark matter, and the stellar luminosity directly correlates with the stellar mass,
which is the dominant component of the baryonic mass in a galaxy. The cold dark matter
models of galaxy formation predict a relation ($M_{baryon} \sim V_{rot}^{\alpha}$)
between the baryonic mass $M_{baryon}$ and the rotation velocity $V_{rot}$, where
$\alpha \sim 3$ for non-baryonic dark matter (e.g., van den Bosch 2000, Navarro \&
Steinmetz 2000) and $\alpha \sim 3.5$ for the collisional baryonic dark matter (Walker
1999).  The stellar component of the baryonic mass can be estimated from the stellar
luminosity, provided, the stellar mass to light ratio ($M^{*}/L$) is known. Bell \& de
Jong (2001) estimated $M^{*}/L$ as a function of color using the spectro-photometric
stellar population synthesis models of galaxy evolution. Observationally, the value of
$\alpha$ is found in the range $3.5 - 4.0$ (e.g., McGaugh et al. 2000, Bell \& de Jong
2001). It is advantageous to study the TF relation using the near-infrared
luminosities due to the following two reasons: 1) the extinction due to dust is
minimal in these bands and, 2) the total stellar mass in galaxies is
dominated by the old stellar population, which emits mostly in the near-infrared.

The I-band TF relation in several groups and clusters has been studied by Giovanelli
et al. (1997) using \HI line-widths and optical rotation curves. They noticed that the
TF relations in the Eridanus and Centaurus groups, and in the Fornax cluster have
larger scatter ($\sigma \sim 0.32-0.45$~mag)  compared to that in other groups and
clusters ($\sigma < 0.3$). One of the implications of the larger scatter could be
significant  distance variations for galaxies in those groups or clusters. Giovanelli
et al. (1997) discussed the selection effects and biases inherent in a sample of
galaxies and their consequences on the slope and on the scatter in the TF relations.
Furthermore, the observed scatter will also depend on the sample size. In addition
to the above effects, various extinction corrections applied to obtain the intrinsic
luminosity of a galaxy are not well understood. The line-width or the rotation
velocity also needs to be corrected for the random and turbulent motions in the
galactic-disks. Tully \& Fouque (1985) have described the procedures to make various
corrections to the two TF observables, rotation velocity and the stellar luminosity.
Many of these corrections are empirically determined. There are indications that the
TF slope depends upon the types of galaxies (Rubin et al. 1985). Persic et al. (1996)
have shown a strong dependence of the shape of the rotation curve on the I-band
luminosity. They noticed that galaxies with M$_{I} < -22.0$ have rotation curves
declining in their outer regions. The dependence of the TF relation on the shapes of
rotation curves has been recently brought to attention by Verheijen (2001). For
galaxies in the Ursa-Major group, they reported a scatter as low as 0.21~mag in the
K-band TF relation, consistent with the measurement uncertainties and leaving no scope
for any intrinsic scatter in the TF relation. They also reported that the scatter in
the near-infrared TF relation is reduced by $\sim0.04$~mag if the flat rotation
velocity is used instead of the maximum rotation velocity. Therefore, it appears that
the flat part of the rotation curve, and hence the dark matter is better correlated
with the stellar luminosity.

The TF relations for the Eridanus group have been studied in the I-band (Giovanelli et
al. 1997) and in the near-infrared bands (Bamford 2002). The I-band  TF slope for the
Eridanus group ($-7.88\pm0.56$) is slightly steeper than the mean slope ($-7.6$) for
all the clusters studied by Giovanelli et al. (1997). Bamford (2002) reported a K-band
TF slope of $-10.8$ and a scatter of $0.58$ for the Eridanus group. 

In this paper, the optical R-band and the near-infrared TF relations are constructed
using the \HI rotation curves of galaxies in the Eridanus group using the GMRT
observations. The \HI observations and the data analyses are described in Omar \&
Dwarakanath (2005). The near-infrared photometry in the J, H, and K bands are from the
{\it Two Micron All Sky Survey} (2MASS). The optical R-band photometric data are from
the 104-cm Sampurnanand telescope at Nainital (Omar 2004).

\section{The Radio observations}

The \HI 21cm-line observations of the Eridanus galaxies were carried out with the
Giant Meterwave Radio Telescope (GMRT). The details of the observations and of the
data analyses can be found in Omar \& Dwarakanath (2005). Here, only a brief
description is given. Galaxies were not selected based on their \HI content or the
optical luminosity. The disk galaxies were selected primarily from the inner 4 Mpc
region of the group where the galaxy density is high. This selection criterion was
adopted since one of the aims of this survey was to study the galaxy evolution in the
high galaxy density regions. After editing and calibrating the data, a channel rms of
$\sim 1$ mJy beam$^{-1}$ was obtained. The velocity resolution was typically $\sim
13.4$ km s$^{-1}$ and the spatial resolution was $\sim20''$ ($\sim 2$~kpc).   

The velocity fields of galaxies were obtained from one-component Gaussian fittings of
the line-profiles at each pixel containing \HI signal. It was found that this
procedure of constructing velocity field provides better image fidelity compared to
that obtained from the moment analysis. The kinematical parameters (position angle,
inclination and rotation curve) of the \HI disks were obtained using the tilted-ring
model. Detailed analyses, maps and results are presented in Omar \& Dwarakanath
(2005). The galaxies often showed kinematical lopsidedness. However, the rotation
curves used in the present analyses are averages of those for the approaching and the
receding sides. The rotation curves of the galaxies in the current study have no
appreciable difference between the maximum and the flat rotation velocities.
Therefore, the TF studies are carried out using the flat part of the rotation curves,
which do not require any significant correction for the effects of beam smearing. The
flat parts of the rotation curves were inferred by visual inspection.

\section{The Optical data}

\subsection{R-band observations and data reduction}

The optical observations were carried out in the R-band ($Cousins$) using the 104-cm
Sampurnanand reflector at the Aryabhatta Research Institute of observational-sciencES
(ARIES, {\em formerly State Observatory}; longitude $79^{o}27'$~E, latitude
$29^{o}22'$~N, altitude $1955$~m), Nainital. This telescope uses the
Ritchey-Chr\'etien system with a f/13 Cassegrain focus having a plate scale of
$15''.5$ mm$^{-1}$. The observations were carried out during October 11 - 17, 2002.
These observations were carried out during a month when the atmospheric conditions at
the site are good. The data were recorded on  $2048\times2048$ pixels CCD camera
cooled at the temperature of liquid Nitrogen. The CCD covers a field of view
$13'\times13'$ on the sky with each square element having a resolution of
$0''.38\times0''.38$. The data were acquired after averaging the counts in every
$2\times2$ pixels making each square element equal to $0''.76\times0''.76$. The
read-out noise and the gain of the CCD camera are 10~e$^{-}$ and 5.3~e$^{-}$ per
analog-to-digital count respectively. Two to five frames each of 5~min or 3~min
duration were observed for each galaxy. A few galaxies having low surface brightness
features were observed for a total of 30~min duration. The observational parameters
are given in Tab.~\ref{tab:Robs}. 

\begin{table}
\begin{center}
\caption{R-band observations}
\label{tab:Robs}
\begin{tabular}{lccccc}
\hline
Name       &Frames &Time &Airmass &FWHM \\
\hline
ESO~482-~G~035		&2	&10	&1.71	&2.8	\\
ESO~548-~G~065		&2	&10	&1.57	&2.6	\\
ESO~549-~G~002		&2	&10	&1.95	&2.8	\\
IC~1952			&3	&10	&1.66	&2.3	\\
IC~1953			&4	&20	&1.87	&2.9	\\
IC~1962			&3	&15	&1.87	&2.4	\\
MCG~-03-10-041		&2	&10	&1.76	&3.1	\\
NGC~1325		&2	&10	&1.66	&2.6	\\
NGC~1371		&3	&15	&1.66	&3.2	\\
NGC~1385 		&6	&30	&1.65	&2.7	\\
NGC~1414		&2	&10	&1.68	&2.9	\\
NGC~1422		&2	&10	&1.68	&2.9	\\
SGC 0321.2-1929		&2	&30	&1.54	&2.0	\\
UGCA~068		&2	&10	&1.77	&3.6	\\
UGCA~077		&2	&10	&1.59	&3.1	\\

\hline

\multicolumn{5}{p{4in}}{\ch $Notes:$ {\bf Column~1:} Name of the galaxy, {\bf Column~2:} Number
of frames, {\bf Column~3:} Total exposure time (in minutes) combining all the frames, {\bf
Column~4:} The mean value of the airmass estimated at the mid-exposure time of the frames; {\bf
Column~5:} The FWHM (in arcsec)of stellar images on CCD estimated after co-adding the frames.}

\end{tabular}
\end{center}
\end{table}

On October 13, 2002 the standard Landolt field SA~92 (Landolt 1992) was observed in
the V and R bands for photometric calibration and  local atmospheric extinction
correction. A total of eight stars from this Landolt field with V magnitudes between
12.5 and 15.6, and, (V--R) colors between 0.31 and 0.72 were observed at different
zenith angles. The value of the zero point is $20.79 \pm 0.03$ and that of the
extinction coefficient ($k$) is  $0.18 \pm 0.02$ on that night. The value of the color
coefficient to make the color correction was insignificant. Hence, no color
corrections were made. The typical value for the R-band sky brightness was 20.5 mag
arcsec$^{-2}$. The average $FWHM$ was $2''.8\pm0''.5$. Based on the vast observing
experience at the observatory, if weather is stable during the  observing period, no
significant variations in the calibration is expected during the nights. The weather
during the observing period was found to be stable. Hence, the Landolt field to
determine the calibration was observed during one of the five nights. The calibration
errors are expected to be less than 0.05~mag, estimated from the fitting errors in the
extinction coefficient and the zero-point. These observations lie in the regime where
noise in the image is dominated by the sky brightness. We estimate, for a typical
galaxy size of $4'$, an additional error of 0.03, 0.15, and 0.7~mag for a 10, 12, and
14~mag galaxy respectively due to Poisson noise.

On each night of observation, several bias frames were taken and averaged to correct
the CCD frames. Since the exposure time in each frame was short, no significant dark
counts are expected. Every alternate night, morning twilight flat fields were
observed. For the nights when no flat fields were taken, previous day's flat fields
were used. The CCD data reduction was performed using the package CCDRED in  the {\it
Image Reduction and Analysis facility} (IRAF) software package developed by {\it
National Optical Astronomy Observatory}.  The frames were aligned using bright stars
in the field. The pixels hit by cosmic rays were removed using the median filtering in
each pixel of different frames whenever at least three frames for a field were
available. When there were only two frames available, the bad pixels were identified
by comparing their values with the average value of the surrounding pixels. After
removing the cosmic ray events, all the aligned frames  for a galaxy were averaged to
get the final image. The images were regrided to a pixel size of $1''$. The images
were registered for the equatorial coordinates in the J2000 epoch using the {\it
Digitized Sky Survey} (DSS) plates.

The sky background was taken as the mode of the pixel values in the averaged frame. If
the presence of a bright star near the edge or within the observed frame caused
significant light gradients across the frame, or, the electron pickup noise caused
bright and dark stripes in the frames, a mean value of the sky was estimated using
pixels surrounding the galaxy.  The Galactic foreground stars over a galaxy were
replaced by an average value of the intensities surrounding the affected region. The
azimuthally averaged limiting surface brightness is $\sim26.0$ mag~arcsec$^{-2}$ in
the R band.

\begin{figure}
\centering
\includegraphics[width=13cm, angle=0]{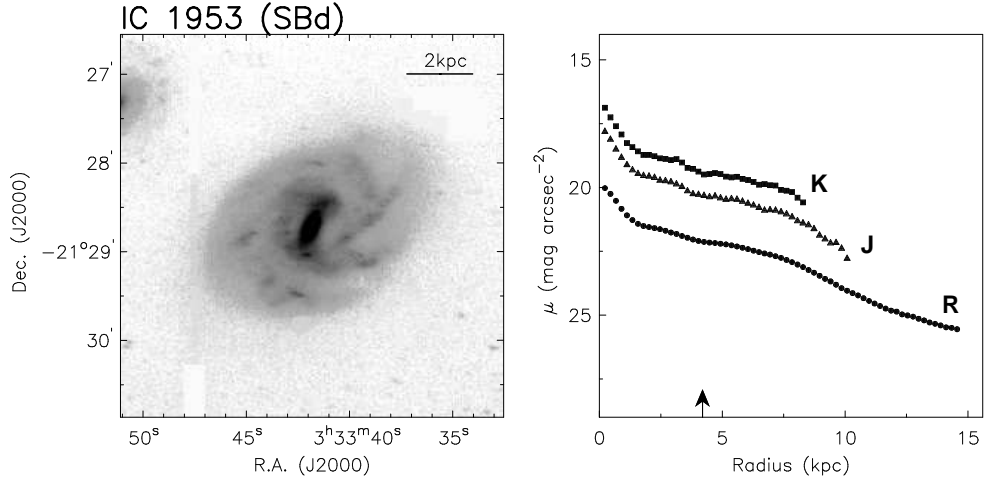}

\caption{(left) The R-band image of IC 1953 obtained using the 104-cm telescope at
Nainital. (right) The average surface brightness profiles in R, J, and K bands as a
function of galacto-centric distance. The J and K band images were obtained from the
2MASS archival data. The arrow on the x-axis marks the disk scale-length in the
R-band.}

\label{fig:Sample}
\end{figure}

The R-band image of IC~1953 and its surface brightness profile are shown in
Fig.~\ref{fig:Sample} as an example. More images can be found in Omar (2004). 

\subsection{2MASS archival data}

The 2MASS is an all sky survey in the near-infrared J (1.24 $\mu$m), H (1.66 $\mu$m)
and K$_{s}$ (2.16 $\mu$m) bands (Jarrett et al. 2000). Each 2MASS camera consists of a
liquid Nitrogen cooled cryostat, housing three $256\times256$ NICMOS3 CCDs each
covering a field of view $8'.5\times8'.5$ on the sky. The final calibrated, co-added
(6.8s integration time), and full resolution images with $1''.0\times1''.0$ sampling
are obtained from the 2MASS archive. 

The overall photometric accuracy of the 2MASS images is believed to be better than
$\sim0.1$~mag (Jarrett et al. 2000). The zero points are typically $20.9$, $20.5$, and
$19.9$ for the J, H, and K bands respectively. The sky brightnesses are typically
$15.3 - 16.0$, $14.3$, $12.9 - 13.5$ mag~arcsec$^{-2}$ in the J, H, and K bands
respectively. The azimuthally averaged surface brightness sensitivities (3$\sigma$)
are $\sim21.0$, $\sim20.5$, and $\sim 20.0$ mag~arcsec$^{-2}$ in  the J, H, and K
bands respectively.

\section{Image analyses}

The image analyses were carried out using IRAF, KARMA, and GIPSY ({\it Groningen Image
Processing System}). All the images were converted to an identical format with $1''.0
\times 1''.0$ pixel size and regrided to the J2000 epoch. Foreground stars on
galaxies, bad pixels, and other defects in the images were removed and replaced by the
mean of the pixels surrounding the affected region. The axial parameters of the
stellar disks (position angle and ellipticity), surface brightness profiles, color
profiles, disk scale lengths, central disk surface brightnesses, isophotal and total
magnitudes etc. were estimated. Details of the analyses and results are described in
Omar (2004). Here, the procedure to obtain the magnitudes of galaxies is described.
The photometric data are presented in Table~5 (Appendix I).

\subsection{Isophotal and total magnitudes}

Both isophotal and total magnitudes were obtained for the galaxies. The isophotal
magnitudes were estimated within the isophote at $26.0$, $21.0$, $20.5$, and $20.0$
mag~arcsec$^{-2}$ for the R, J, H, and K bands respectively. These are the limiting
isophotes for the respective bands. The isophotal magnitude $m_{l}^{iso}$ corrected
for the atmospheric extinction is estimated using the following relation:

\begin{equation}
m_{l}^{iso} = Z - 2.5log(S_{l}/t) - k.X
\end{equation}

\noindent where the subscript $l$ refers to the limiting isophote in a band and the
superscript $'iso'$ refers to isophotal, $Z$ is the zero point magnitude corresponding
to 1 second of exposure time, $k$ is the extinction coefficient, $X$ is the airmass,
$S_{l}$ is the total number of counts above the sky background within the limiting
isophote, and $t$ is the total exposure time on the galaxy.  

The isophotal magnitude provides a lower limit to the actual luminosity of a galaxy.
Normally, optical observations are sensitive to detect light up to $4 - 5$ disk scale
lengths while  near-infrared observations detect light only up to $2 - 3$ disk scale
lengths in a reasonable exposure time. It is obvious that different bands are
sensitive to different levels in detecting light from galaxies. To overcome this
effect, correction for the un-detected light has to be applied to obtain the total
magnitudes of galaxies. If the disks were infinitely extended and follow the
exponential light profiles, it can be shown that the total light above a certain
limiting isophotal radius depends only on the number of scale lengths within that
limiting radius, and can be estimated using the following relation : 

\begin{equation}
\delta m_{ext} = 2.5log\left(1 - (1 + \delta n)exp(-\delta n)\right)
\label{eqn:cormag}
\end{equation}

\noindent (Tully et al. 1996); where $\delta m_{ext}$ is the total light beyond the
limiting isophote radius and $\delta$n is the number of observed scale lengths within
the limiting isophote radius $\left({\delta}n = (\mu_{0}-\mu_{lim})/1.086 \right)$.
Here, $\mu_{0}$ and $\mu_{lim}$ are the central disk surface brightness and the
surface brightness of the limiting isophote respectively. It is suggested by Tully et
al. (1996) that the disks are in reality not infinitely extended but rather truncated.
They empirically determined that only 88\% of the $\delta m_{ext}$ given by
Eq.~\ref{eqn:cormag} should be used to make correction to the isophotal magnitude in
order to get the total magnitude.

The scale length $r_{d}$, and the central disk surface brightness $\mu_{0}$ were
estimated from a fit of the form $I(r) = I(0) e^{-r/r_{d}}$ to the averaged radial
surface brightness profile (see Omar 2004 for details). The average radial surface
brightness ($I(r)$) was computed for galaxies in elliptical annuli of width $2''$ with
a fixed position angle and ellipticity. The inner regions where the disk profiles are
significantly modified due to the presence of a bar or a bulge were excluded in the
fit. The range of data to be excluded in the fit was determined by visual inspection.

\subsection{Extinction corrections}

The total magnitudes were corrected for the Galactic and the internal dust extinction
in galaxies. The values of the Galactic extinction provided by Schlegel et al (1998)
were used in the present analysis.  The high Galactic latitude ($\sim-52\deg$) of the
Eridanus group results in minimal Galactic extinction. The average values of the
Galactic extinction toward Eridanus are $\sim0.11$, $\sim0.04$, $\sim0.02$, and
$\sim0.01$~mag in the R, J, H, and K bands respectively. The estimation of the
internal dust extinction in galaxies is quite uncertain. It depends on their
orientation, morphological type, size, and luminosity. Several authors have discussed
internal dust extinction in galaxies (e.g., Tully \& Fouque 1985, Tully et al. 1998,
Masters et al. 2003).  We used the corrections suggested by Tully et al. (1998) and
Masters et al. (2003), who obtained an empirical relation between the total luminosity
and the dust content in disk galaxies. The extinction, $A_{\lambda}^{i}$, is given as
a function of the axial ratio ($b/a$) in the pass-band $\lambda$ as:

\begin{equation} 
A_{\lambda}^{i} = \gamma_{\lambda}~ log~(a/b) 
\end{equation}

\noindent Where $a$ and $b$ are the length of the semi-major and the semi-minor axis
respectively. The $\gamma$ was estimated in each band using an equation of the
following form:

\begin{equation}
\gamma_{\lambda} = c_{0} + c_{1}(M_{\lambda} + c_{2})
\end{equation}

\noindent where the values of the coefficients $c_{i}$ (i=0,1,2) for each band and for
different magnitude ranges are given in Tab.~\ref{tab:gamma}. $M_{\lambda}$ is the
absolute magnitude in the corresponding pass-band.

\begin{table}
\begin{center}
\caption{Coefficients for internal dust extinction correction.}
\label{tab:gamma}
\begin{tabular}{llccc}
\hline
Band & mag. range & $c_{0}$ & $c_{1}$ & $c_{2}$ \\
\hline 

R & $> -16.2$ & 0.00 & 0.00 & 0.00 \\
  & $< -16.2$ & 0.00 &-0.24 & 16.2 \\
J & $> -22.2$ & 0.77 &-0.26 & 23.0 \\
  & $< -22.2$ & 0.60 &-0.04 & 23.0 \\
H & $> -22.2$ & 0.58 &-0.26 & 23.0 \\
  & $< -22.2$ & 0.38 &-0.01 & 23.0 \\
K & $> -22.2$ & 0.31 &-0.13 & 23.0 \\
  & $< -22.2$ & 0.20 & 0.00 & 23.0 \\
\hline
\end{tabular} 
\end{center} 
\end{table}

\section{Sample selection and the error budget}

The slope and the scatter in the TF relation depends on the selection criteria and
biases  in the sample. Bernstein et al. (1994) outlined some criteria for galaxies to
be included in TF studies. They preferred non-interacting galaxies of types Sb -- Sd
with steep \HI profiles at the end velocities, having smooth outer isophotes and
without a prominent bar. Galaxies with inclination $<45\deg$ are generally not
included in TF studies. This is because for low inclination galaxies, it is difficult
to get an accurate estimate of the position angle and the inclination angle. This
results in uncertain deprojected rotation velocities. For instance, for a TF slope of
$-10$ in the near-infrared, a marginal error of 5\deg in the inclination angle
introduces an additional scatter of nearly 0.4~mag, 0.2~mag and 0.05~mag in the TF
relations at inclinations of 45\deg, 60\deg, and 80\deg respectively. At the same
time, nearly edge-on galaxies will have their optical luminosities poorly estimated
due to uncertainties in the internal dust extinction corrections. Therefore, TF
studies on galaxies having intermediate inclination angles, and in the near-infrared
bands are preferred. One would also like to reduce scatter due to uncertainties in the
distances to galaxies in a sample. For this reason, galaxies in a group or a cluster
may be the right choice instead of field galaxies. 

The survey of the Eridanus group was not volume-limited. The rotation curves were measured
for about half of the spiral galaxies. Therefore, somewhat relaxed criteria
compared to those described above are used to construct the sample for the present
study. The S0/a and interacting galaxies were excluded from the sample. Galaxies with
inclination less than $35\deg$ are not included. There are 4 galaxies (ESO 548- G 065,
ESO 549- G 002, IC 1962, NGC 1422), whose rotation curves are nearly flat. The sample
consists of finally a total of 17 galaxies for which the \HI rotation curves are flat
or nearly flat in the outer regions. These galaxies are included in the sample. The
photometric data on all galaxies are not available in all the bands. The photometric
and the kinematical properties of the galaxies in the Eridanus group are given in
Tab.~\ref{tab:prop}. 

We have assumed all the galaxies to be at a distance of  $\sim23\pm2$~Mpc. This assumption
is based upon the distance estimates using the surface brightness fluctuations in the
I-band and in the K-band of a few early type galaxies in the different sub-groups of
the Eridanus group (Jensen et al. 1998, Tonry et al. 1997, Tonry et al. 2001). The
mean distance modulus $(m-M)$ of NGC~1400 (S0, 558~\km), NGC~1407 (E, 1779~\km),
NGC~1395 (E, 1717~\km), NGC~1332 (S0, 1524~\km), and NGC~1426 (E, 1443~\km)  is
$31.8\pm0.2$~mag.

\begin{table}
\caption{Sample of galaxies used to construct the TF relation.}
\label{tab:prop}
\centering
\begin{tabular}{lccccccc}
\hline
Name & Type &Incl. &log(W) &M$_{R}$ &M$_{J}$ &M$_{H}$ &M$_{K}$  \\
\hline

ESO 482- G 005 & SBdm & 84.6 & 2.205 & -- & -17.31 & -- & --  \\
ESO 482- G 035 & SBab & 51.8 & 2.374 & -18.86 & -20.62 & -21.08 & -21.45  \\
ESO 548- G 021 & SBdm  & 88 & 2.244 & -- & -18.47 & -18.19 & -18.99 \\
ESO 548- G 065 & SBa & 79.6 & 2.164 & -16.96 & -18.29 & -- & --  \\
ESO 549- G 002 & IBm & 53.1 & 2.048 & -17.46 & -18.33 & -- & --  \\
IC 1952 & SBbc & 81.4 & 2.427 & -19.39 & -21.55 & -22.09 & -22.19  \\
IC 1953 & SBd & 50.6 & 2.476 & -20.18  & -22.13 & -22.68 & -23.04  \\
IC 1962 & SBdm & 79.8 & 2.216 & -17.64 & -18.24 & -18.21 & -19.41  \\
MCG -03-10-041 & SBdm & 62.5 & 2.342 & -18.25 & -19.92 & -20.44 & -20.54  \\
NGC 1325 & SAbc & 74.5 & 2.500 & -20.69 & -22.64 & -23.18 & -23.33 \\
NGC 1371 & SABa & 48.1 & 2.716 & -21.12 & -23.38 & -24.00 & -24.21  \\
NGC 1385 & SBcd & 51.5 & 2.442 & -20.97 & -22.80 & -23.41 & -23.55  \\
NGC 1414 & SBbc & 79.8 & 2.193 & -16.95 & -19.27 & -19.68 & -20.10  \\
NGC 1422 & SBab & 79.6 & 2.169 & -17.95 & -20.38 & -20.95 & -21.16  \\
SGC 0321.2-1929 & IBm & 36.2 & 2.052 & -15.04 & -- & -- & --  \\
UGCA 068 & SABcdm & 37.6 & 2.244  & -17.57 & -18.90 & -19.15 & -19.65  \\
UGCA 077 & SBdm & 64.3 & 2.193 & -17.30 & -- & -- & --  \\
\hline

\multicolumn{8}{p{5in}}{\ch Notes: {\bf Column~1:} Name of the galaxy, {\bf Column~2:}
Hubble type. {\bf Column~3:} The inclination angle (in degree) as obtained from the
tilted ring model or obtained from the R-band optical images in those cases where the
tilted ring model did not provide a satisfactory fit, {\bf Column~4:} Width $W$ is
twice the value of the velocity at the flat part of the rotation curve, or, of the
rotation velocity at the last measured point. {\bf Columns~5-8:} Absolute total
magnitudes in the R, J, H, and K bands respectively. These magnitudes include all the
corrections as described in Sect.~4. All the galaxies are assumed to be at a distance
of 23~Mpc (distance modulus $\sim -31.8$).}

\end{tabular} 
\end{table}

The photometric precision is estimated taking into account the measurement
uncertainties. For typical magnitudes of the Eridanus galaxies ($m\sim12$~mag) in the
R-band, the photometric accuracy is estimated as $\sim0.2$~mag (Omar 2004). The
accuracy of the 2MASS photometry is taken as 0.1~mag. There seems to be not much
understanding on how much additional scatter can arise in the TF relations due to
uncertainties in the internal dust extinction. Verheijen (2001) noted that the scatter
in the TF relation is increased by a marginal amount ($\sim0.04$~mag) from the K band
to the R-band. If we attribute this extra scatter to the extinction corrections, it
appears that the internal dust extinction corrections are fairly well determined in
these bands.  There may be a significant contribution to the total scatter due to
errors in the inclination angles of galaxies as explained before. We estimate the
error budget for an intermediate inclination of 60\deg. Though all the errors may not
be truly random in nature, we estimate the total expected rms assuming that the
different errors add up randomly. This combined error is estimated as $0.3$~mag.

\section{Tully-Fisher relations}

\subsection{$Classical$ luminosity-linewidth relation}

\begin{figure}
\centering
\includegraphics[width=13cm, angle=0]{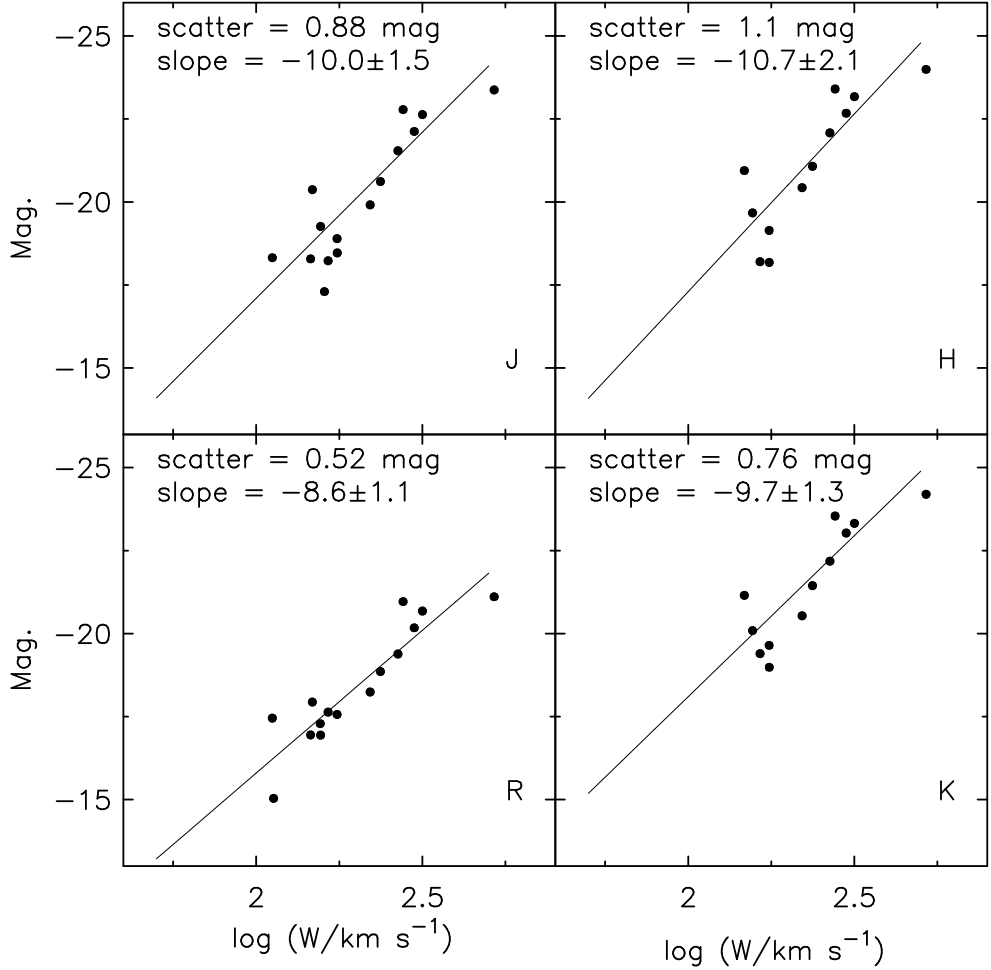}

\caption{The Tully-Fisher relations in the  R, J, H, and K bands. The values of log W
(abscissa) and absolute magnitude (ordinate) are from Tab.~\ref{tab:prop}.}

\label{fig:TFR}
\end{figure}

Figure~\ref{fig:TFR} shows TF relations in the R, J, H, and K bands. A least square
fit is carried out to obtain the slope $'a'$ and the intercept $'b'$ for an equation
of the form $Mag. = a~ log(W) + b$. The fit is carried out using data for galaxies
having flat rotation curves ($W = 2V_{flat}$). All such data points are given equal
weights. The total observed scatter (rms of residuals) is estimated in magnitude. The
slope, the intercept, and the scatter obtained from the fit are given in
Tab.~\ref{tab:TFRparam}. 

The slopes of the TF relations obtained in the present analysis are comparable to
those obtained in other studies of nearby groups and clusters (e.g., Giovanelli et al.
1997, Verheijen 2001). The TF slopes from the current study are  $-8.6\pm1.1$,
$-10.0\pm1.5$, $-10.7\pm2.1$, $-9.7\pm1.3$ in the R, J, H, and K bands respectively.
The slopes obtained by Verheijen (2001) for the Ursa-Major galaxies are $-8.3\pm0.6$ 
and $-11.2\pm0.7$ in the R and K bands respectively. They reported a scatter as low as
$0.21$ mag for the K-band. Giovanelli et al. (1997) obtained a mean slope of $-7.6$ in
the I-band for a set of nearby groups and clusters. They also reported an I-band TF
slope of $-7.88\pm0.56$ with a scatter of $0.36$~mag for the Eridanus group using
their single dish \HI data and the optical rotation curves. They pointed out that both
the slope and the scatter for the Eridanus group are somewhat different from the
average values for other groups and clusters in their sample. Recently, using  the
photometric data from  2MASS and the kinematical data from Giovanelli et al. (1997), 
Bamford (2002) reported slopes of $-10.8$, $-10.9$, and $-11.1$ for the J, H, and K
bands respectively, and a scatter of $0.57$ mag for the Eridanus galaxies. It can be
seen from Tab.~\ref{tab:TFRparam} that the scatter in the TF relations in the present
study ($0.5 - 1.1$ mag) is also significantly higher compared to that obtained for the
Ursa major group and for other nearby groups and clusters.

The larger scatter in the TF relations for the Eridanus galaxies in various studies
including those based on the near-infrared data indicate that there are variations in
the distances to the galaxies in the group. If the additional scatter in the TF
relation is attributed to the depth of the group then it indicates that the depth of
the Eridanus group can be as large as 4~Mpc. Perhaps, it is tied up with the dynamical
stage of the Eridanus group where galaxies or small groups are being accreted from
outer regions (Willmer et al. 1989). 

\subsection{Baryonic TF relation}

\begin{figure}
\centering
\includegraphics[width=13cm, angle=0]{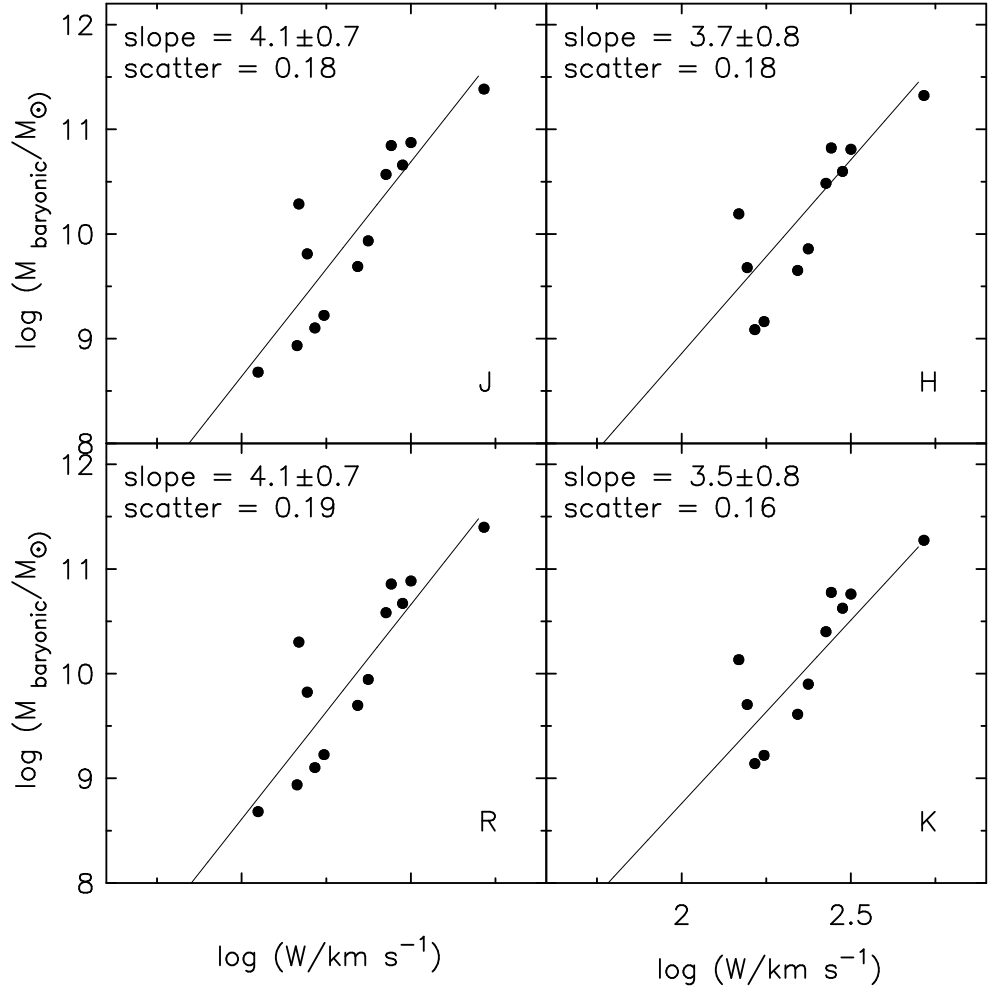}

\caption{The baryonic TF relations constructed using the combined \HI and Helium gas
mass, and the stellar mass obtained using the stellar luminosities in the R, J, H, and
K bands and their corresponding M$^{*}$/L ratios from the models of Bell \& de Jong
(2001).}

\label{fig:baryTF}
\end{figure}

The baryonic TF relations are constructed using the combined \HI and Helium gas mass,
and the stellar mass obtained using the stellar luminosities in different bands and
their corresponding M$^{*}$/L ratios from the spectrophotometric population synthesis
models of Bell \& de Jong (2001). These models predict a linear relationship between
the colors of galaxies and the M$^{*}$/L  ratio for a band. These models allow the
estimation of the total stellar mass from the stellar luminosity in any wave-band
provided the colors of galaxies are known.  For the current study, the total stellar
masses are estimated using the (R--J) colors. The Helium mass is taken as 1.32 times
the \HI mass. Since the molecular and the ionized gas masses are  unknown, the
baryonic masses thus obtained are lower limits. The baryonic TF relations are plotted
in Fig.~\ref{fig:baryTF}.

If the TF relations were truly a correlation between the baryonic mass and the
rotation velocity, and the models of  Bell \& de Jong (2001) make accurate predictions
of M$^{*}$/L for each band, the baryonic TF slope should be independent of pass-band. 
A nearly pass-band independent baryonic TF relation is obtained in the R, J, H, and K
bands as indicated in Fig.~\ref{fig:baryTF} and Tab.~\ref{tab:TFRparam}. The slope of
the Baryonic TF relation is in the range $3.5 - 4.1$. 

McGaugh et al. (2000) obtained a slope of $3.98\pm0.12$ for a sample of galaxies with
rotation velocities over a large range ($30-300$~\km). Using data from Verheijen
(2001),  Bell \& de Jong (2001) obtained a slope of $3.5\pm0.4$ for the Ursa-major
group of galaxies. The slope of the baryonic TF relation is predicted to be $3$ for
non-baryonic dark matter (Navarro \& Steinmetz 2000) and $3.5$ for the collisional
baryonic dark matter (Walker 1999).  It is worth mentioning here that all these
observational baryonic TF relation (including the present one) do not include the
molecular and the ionized gas masses. These mass components will be significant for
dwarfs and gas rich low surface brightness (LSB) galaxies, whose rotation velocities
are lower compared (typically $< 100 ~km s^{-1}$) to that for high surface brightness
(HSB) galaxies. The inclusion of the molecular and the ionized gas masses to the total
baryonic mass in a sample having all types (dwarfs, LSB and HSB) of galaxies will make
the slope of the baryonic TF relations shallow. Further, the M$^{*}$/L ratios in the
models of Bell \& de Jong (2001) were calibrated assuming that the galactic disks are
supported mainly by rotation. If this is not the case, the baryonic TF relation will
be even shallower. We believe that with uncertainties in these analyses and the
incomplete information about the total baryonic mass in galaxies, currently no dark
matter model can be ruled out based on the baryonic TF relation.

\begin{table}
\begin{center}
\caption{The TF parameters.}
\label{tab:TFRparam}
\begin{tabular}{lccc}
\hline
Band  & Slope & Intercept & Scatter \\
\hline
\multicolumn{4}{p{3in}}{\bf Classical TF} \\
& & &   \\
R & $-8.6\pm1.1$   & $1.4\pm2.5$  & 0.52 mag \\ 
J & $-10.0\pm1.5$  & $2.9\pm3.5$  & 0.88 mag \\
H & $-10.7\pm2.1$  & $4.1\pm5.0$  & 1.10 mag \\
K & $-9.7\pm1.3$   & $1.3\pm4.1$  & 0.76 mag \\
 & & &  \\
 
\hline

\multicolumn{4}{p{3in}}{\bf Baryonic TF} \\
 & & & \\
R & $4.1\pm0.7$ & $0.4\pm1.6$  & 0.19 \\
J & $4.1\pm0.7$ & $0.4\pm1.6$  & 0.18 \\
H & $3.7\pm0.8$ & $1.5\pm2.0$  & 0.18 \\
K & $3.5\pm0.8$ & $1.8\pm1.8$  & 0.16 \\
\hline

\multicolumn{4}{p{3in}}{Note: The scatter is the reduced chi-square value of the
fit.}

\end{tabular} 
\end{center} 
\end{table}

\section{Conclusions}

\begin{itemize}

\item The TF slopes in the Eridanus galaxies are consistent with those obtained in
other studies of nearby groups and clusters. The slopes of the TF relations (absolute
magnitude $vs$ log$2$V$_{flat}$) for the galaxies in the Eridanus group are
$-8.6\pm1.1$, $-10.0\pm1.5$, $-10.7\pm2.1$, and $-9.7\pm1.3$ in the R, J, H, and K
bands respectively.

\item The TF relation for the Eridanus group has a significantly larger scatter
($\sigma \sim 0.5-1.1$~mag) compared to that for other groups and clusters ($\sigma
\sim 0.2-0.4$~mag).  

\item If the larger scatter in the TF relations indicates variations in the
distances to  galaxies in a sample, the maximum difference in the distances to
galaxies in the Eridanus group will be $\sim4$~Mpc. This implies that the galaxies in
the Eridanus group are more dispersed in space compared to other groups and clusters. 
It is suspected that we arr looking down a filament.

\item The value of the index $\alpha$ ($M_{baryon} \sim V_{flat}^{\alpha}$) of the
baryonic TF relation is in the range $3.5 - 4.1$.

\end{itemize}

\section*{Acknowledgments}

We thank Shashi Pandey and Ram Sagar for the help provided during the optical
observations from Nainital. 2MASS data obtained as part of the Two Micron All Sky
Survey (2MASS), a joint project of the University of Massachusetts and the Infrared
Processing and Analysis Center/California Institute of Technology, funded by the
National Aeronautics and Space Administration and the National Science Foundation. 

\newpage

\section*{Appendix-I}

\begin{table}
\begin{center}
\caption{Photometric Results}
\label{tab:app}
\begin{tabular}{llccccccc}
\hline
\hline
Galaxy        &Type          &F   &r$_{d}~~~~~~\pm$ &r$_{(25)}$    &$\mu^{0}~~~~~~\mu^{0,i}~~~~\pm$   &m$_{26}$  &m$^{total}$ \\
              &P.A.$\pm$     &    &                 &r$_{(21)}$    &                                  &m$_{21}$  &            \\
              &Incl.$\pm$    &    &                 &r$_{(20.5)}$  &                                  &m$_{20.5}$&            \\
	      &r(B$_{25}$)   &    &                 &r$_{(20)}$    &                                  &m$_{20}$  &            \\
              &(kpc)         &    &(kpc)            &(kpc)         &(mag arcsec$^{-2}$)                   &(mag)     &(mag)       \\
\hline
E~482-005 &SBdm  &R & -- &--&-- &--\\
	  &79~~0 &J &1.89~~0.10 &2.5 &19.51~~22.06~~0.05 &15.10 &14.50 \\
	  &85~~2 &H &-- & --  &-- & --&-- \\
	  &5.6   &K &-- & -- &-- &-- & --\\
          &      &  &           &    &                   &      &      \\
E~482-035 &SBab  &R &2.20~~0.03 &7.4 &21.02~~21.23~~0.02 &13.19 &12.94 \\
          &4~~~1  &J &1.95~~0.05 &3.6 &18.89~~19.39~~0.04 &11.76 &11.18 \\
          &52~~1 &H &1.61~~0.08 &3.6 &18.13~~18.64~~0.08 &11.18 &10.72 \\
          &4.4   &K &1.96~~0.12 &3.4 &18.14~~18.65~~0.09 &11.03 &10.35 \\
          &      &  &           &    &                   &      &      \\
E~548-021 &SBdm  &R &-- & --  & --&-- &-- \\
          &70~~1 &J &4.71~~0.20 &5.6 &19.40~~23.01~~0.02 &14.12 &13.33 \\
          &88~~0 &H &3.37~~0.23 &4.5 &18.86~~22.48~~0.05 &14.15 &13.62 \\
          &6.7   &K &3.71~~0.34 &4.5 &18.43~~22.06~~0.06 &13.68 &12.81 \\
          &      &  &           &    &                   &      &      \\
E~548-065 &SBa   &R &1.77~~0.01 &5.8 &21.34~~22.11~~0.01 &15.31 &14.85 \\
          &41~~0 &J &1.74~~0.10 &2.2 &18.99~~20.78~~0.05 &13.98 &13.51 \\
          &80~~1 &H &  --    & --  &   --            &   --  &  --   \\
          &5.0   &K &  --     & --  &      --         &  --  &  --  \\
          &      &  &           &    &                   &      &      \\
E~549-002 &IBm   &R &1.51~~0.03 &4.5 &21.79~~21.88~~0.03 &14.73 &14.34 \\
          &32~~1 &J &1.08~~0.15 &1.3 &19.42~~19.91~~0.21 &14.28 &13.48 \\
          &53~~1 &H &  --     & --  &     --          &  --   &   --  \\
          &4.4   &K &   --    & --  &      --        &  --   &  --   \\
	  &      &  &           &    &                   &      &      \\
I~1952    &SBbc  &R &3.52~~0.06 &11.2 &20.21~~21.21~~0.02 &13.05 &12.42 \\
          &140~0&J &3.60~~0.04 &9.0 &17.88~~19.91~~0.01 &10.78 &10.26 \\
          &81~~0 &H &3.46~~0.06 &9.0 &17.17~~19.22~~0.02 &10.15 &9.72 \\
          &8.7   &K &3.49~~0.07 &7.8 &16.92~~18.97~~0.02 &9.98  &9.62 \\
          &      &  &           &    &                   &      &      \\
I~1953    &SBd   &R &4.20~~0.11 &12.1 &21.08~~21.26~~0.03 &11.95 &11.62 \\
          &121~0&J &3.78~~0.10 &7.4  &19.03~~19.50~~0.03 &10.39 &9.68 \\
          &51~~1 &H &3.78~~0.07 &7.6  &18.42~~18.90~~0.02 &9.75 &9.13 \\
          &9.4   &K &4.19~~0.13 &7.2  &18.22~~18.70~~0.03 &9.51 &8.77 \\
\end{tabular}
\end{center}
\end{table}

\begin{table}
\begin{center}
\begin{tabular}{llccccccc}

I~1962    &SBdm  &R &2.77~~0.05 &9.2  &21.64~~22.54~~0.02 &14.62 &14.16 \\
          &1~~~1  &J &2.06~~0.14 &2.5  &19.70~~21.56~~0.07 &14.45 &13.57 \\
          &80~~2 &H &1.76~~0.21 &2.0  &19.29~~21.16~~0.14 &14.44 &13.60 \\
          &9.1   &K &1.68~~0.18 &2.7  &18.32~~20.19~~0.13 &13.19 &12.40 \\
          &      &  &           &    &                   &      &      \\
M-03-10-041 &SBdm   &R &2.67~~0.08 &6.9 &21.81~~22.09~~0.04 &13.99 &13.56 \\
            &166~1 &J &3.13~~0.28 &2.0 &20.14~~20.93~~0.10 &13.54 &11.89 \\
            &63~~1  &H &2.68~~0.26 &2.7 &19.36~~20.17~~0.12 &12.59 &11.37 \\
            &6.7    &K &2.36~~0.29 &2.0 &18.95~~19.77~~0.16 &12.63 &11.26 \\
          &      &  &           &    &                   &      &      \\
N~1325    &SAbc &R &4.93~~0.03  &19.3 &20.45~~21.14~~0.01 &11.75 &11.12 \\
          &53~~1&J &4.60~~0.06  &11.7 &18.04~~19.46~~0.01 &9.74 &9.17 \\
          &75~~0&H &4.51~~0.09  &12.3 &17.36~~18.78~~0.03 &9.06 &8.63 \\
          &15.8 &K &4.61~~0.09  &10.8 &17.13~~18.56~~0.02 &8.88 &8.48 \\
          &      &  &           &    &                   &      &      \\
N~1371    &SABa  &R &3.55~~0.02  &17.2 &20.22~~20.38~~0.01 &10.98 &10.69 \\
          &134~1&J &3.34~~0.02  &10.3 &17.65~~18.07~~0.01 &8.75  &8.42 \\
          &48~~1 &H &3.28~~0.02  &10.8 &16.96~~17.38~~0.01 &8.06  &7.81 \\
          &18.8  &K &3.37~~0.04  &9.6  &16.76~~17.19~~0.01 &7.87  &7.60 \\
          &      &  &           &    &                   &      &      \\
N~1385    &SBcd  &R &3.32~~0.04  &15.2 &20.20~~20.42~~0.02 &11.14 &10.84 \\
          &173~1&J &3.21~~0.10  &8.7  &18.10~~18.60~~0.05 &9.43  &9.02 \\
          &51~~1 &H &3.16~~0.10  &9.2  &17.39~~17.89~~0.05 &8.72  &8.40 \\
          &11.4  &K &2.98~~0.11  &8.1  &17.13~~17.64~~0.07 &8.60  &8.26 \\
          &      &  &           &    &                   &      &      \\
N~1414    &SBbc  &R &1.88~~0.04  &5.6  &21.16~~22.07~~0.02 &15.18 &14.86 \\
          &170~1&J &1.96~~0.06  &3.8  &18.69~~20.55~~0.03 &12.95 &12.54 \\
          &80~~1 &H &1.57~~0.04  &4.0  &17.85~~19.72~~0.04 &12.37 &12.13 \\
          &5.8   &K &1.81~~0.07  &3.4  &17.73~~19.60~~0.04 &12.23 &11.71 \\
          &      &  &           &    &                   &      &      \\
N~1422    &SBab  &R &2.32~~0.03  &9.2  &20.90~~21.80~~0.02 &14.30 &13.86 \\
          &66~~0 &J &2.44~~0.03  &5.8  &18.29~~20.12~~0.01 &11.87 &11.43 \\
          &80~~1 &H &2.12~~0.05  &5.8  &17.38~~19.22~~0.03 &11.15 &10.85 \\
          &7.4   &K &1.84~~0.03  &5.4  &16.94~~18.79~~0.02 &10.98 &10.65 \\
          &      &  &           &    &                   &      &      \\
S~0321.2-1929 &IBm   &R &2.10~~0.20 &0.2 &24.79~~24.83~~0.065 &17.99 &16.76 \\
              &170~0 &J &-- & --&-- & --&-- \\
              &36~~6 &H & --&-- &-- &-- &-- \\
              &4.4   &K & --&-- & --& --& --\\
\end{tabular}
\end{center}
\end{table}

\begin{table}
\begin{center}
\begin{tabular}{llcccccc}

U~068     &SABcdm&R &2.06~~0.05  &5.2  &22.41~~22.45~~0.03 &14.53 &14.24 \\
          &42~~1 &J &1.29~~0.12  &1.3  &19.58~~19.80~~0.13 &13.84 &12.90 \\
          &38~~3 &H &1.43~~0.22  &1.3  &19.27~~19.51~~0.20 &13.75 &12.66 \\
          &5.7   &K &1.08~~0.16  &1.3  &18.55~~18.79~~0.24 &13.08 &12.15 \\
          &      &  &           &    &                   &      &      \\
U~077     &SBdm  &R &2.42~~0.05  &6.0  &22.52~~22.83~~0.02 &14.97 &14.51 \\
          &138~1&J &-- & --  &-- & --&-- \\
          &64~~1 &H & --& -- & --& --& --\\
          &5.7   &K & --& --  & --& --& --\\
\hline
\multicolumn{8}{p{5.5in}}{\ch Notes: 
{\bf Column~1:}  The name of the galaxy in an abbreviated format. The ESO galaxies are written as
E, NGC is written as N, UGCA is written as U, IC is written as I, MCG is written as M, and SGC is
written as S. {\bf Column~2:} The morphological type, position angle and inclination with
associated errors in the estimation determined from the R band observations or from DSS images when
R band observations were not available, radius in kpc at 25 mag arcsec$^{-2}$ B band isophote taken
from RC3. The P.A. is measured from North to East (counter-clockwise). {\bf Column~3:} The
filter/band.  {\bf Column~ 4\&5:} The disk scale length and associated errors in the fitting
procedure in units of kpc. {\bf Column~6:} The radius in kpc corresponding to the isophotal level
of 25 mag arcsec$^{-2}$ in the R, 21 mag arcsec$^{-2}$ in the J, 20.5 mag arcsec$^{-2}$ in the H,
and 20 mag arcsec$^{-2}$ in the K band.  {\bf Column~7,8,\&9:} The central surface brightness of
the disk as obtained from the fit, and the values corrected for the inclination, Galactic
extinction, and internal extinction. The error is obtained from the fitting procedure and does not
include calibration errors. {\bf Column~10:} The isophotal magnitude of the galaxy estimated within
the fixed aperture at 26.0 mag arcsec$^{-2}$ for the R-band,  $21.0$ mag~arcsec$^{-2}$ in the J
band, $20.5$ mag~arcsec$^{-2}$ in the H band, and $20.0$ mag~arcsec$^{-2}$ in the K band. {\bf
Column~11:} The total magnitude extrapolated to infinity with corrections for Galactic and internal
extinctions.}

\end{tabular}
\end{center}
\end{table}


\newpage

\begin{thebibliography}{}
\bibitem{bal74} Balkowski, C., Bottinelli, L., Chamaraux, P., Gouguenheim, L.,
\& Heidmann, J. 1974, {\it Astron. Astrophys.}, {\bf 34}, 43
\bibitem{bamford} Bamford, S.P. 2002, {\it astro-ph/0210227}
\bibitem{bell} Bell, E.F., \& De Jong, R.S. 2001, {\it Astrophys. J.}, {\bf 550}, 212
\bibitem{bern} Bernstein, G.M., Guhathakurta, P., \& Raychaudhury, S. et al. 1994, {\it Astron. J.}, {\bf 107}, 1962
\bibitem{Gio} Giovanelli, R., Haynes, M. P., Herter, T., Vogt, N. P., da Costa, L. N.,
Freudling, W., Salzer, J. J., Wegner, G. 1997, {\it Astron. J. }, {\bf 113}, 53
\bibitem{jarret} Jarrett, T. H., Chester, T., Cutri, R., Schneider, S., Skrutskie, M., \& Huchra, J. P. 2000, {\it Astron. J. }, {\bf 119}, 2498
\bibitem{lando} Landolt A.U. 1992, {\it Astron. J.}. {\bf 104}, 340
\bibitem{mast} Masters K.L., Giovanelli, R., \& Haynes, M.P. 2003, {\it Astron. J. }, {\bf 126}, 158
\bibitem{mcgau} McGaugh, S. S., Schombert, J. M., Bothun, G. D., \& de Blok, W. J.
G.,2000, {\it Astrophys. J. Letter}, {\bf 533}, 99
\bibitem{nav00} Navarro, J.F., \& Steinmetz, M. 2000, {\it Astrophys. J.}, {\bf 538}, 477
\bibitem{omar1} Omar A. 2004, Ph.D. thesis, Jawaharlal Nehru University, Delhi.
\bibitem{omar2} Omar, A. \& Dwarakanath, K.S. 2005, {\it J. Astrophys. Astron.}, {\bf 26}, 1  
\bibitem{persic} Persic, M., Salucci, O., \&  Stel, F. 1996, {\it Mon. Not. R. Astron. Soc.}, {\bf 281}, 27
\bibitem{rob69} Roberts, M.S. 1969, {\it Astron. J. }, {\bf 74}, 859
\bibitem{rog72} Rogstad, , D.H., \& Shostak, G. S. 1972, {\it Astrophys. J.}, {\bf 176},
315
\bibitem{Rubin} Rubin, V. C., Burstein, D., Ford, W. K., Jr.; Thonnard, N. 1985, {\it
Astrophys. J.}, {\bf 289}, 81
\bibitem{sch} Schlegel, D.J., Finkbeiner, D.P., \& Davis, M. 1998 {\it Astrophys. J.}, {\bf 500}, 525
\bibitem{shso75} Shostak, G. S. 1975, {\it Astrophys. J.}, {\bf 198}, 527 
\bibitem{TF} Tully, R. B., \& Fisher, J. R. 1977, {\it Astron. Astrophys.}, {\bf 54}, 661 
\bibitem{Tfo} Tully, R.B., \& Fouque, P. 1985, {\it Astrophys. J. Suppl.}, {\bf 58}, 67
\bibitem{t98} Tully, R.B., Pierce, M.J., Huang, Jia-Sheng et al. 1998 {\it Astron. J.}, {\bf 115}, 2264
\bibitem{t96} Tully, R.B., Verheijen, M.A.W., Pierce, M.J. et al. 1996 {\it Astron. J.}, {\bf 112}, 2471
\bibitem{bosch} van den Bosch, F.C. 2000, {\it Astrophys. J.}, {\bf 530}, 177  
\bibitem{verh} Verheijen, M.A.W. 2001, {\it Astrophys. J.}, {\bf 563}, 694 
\bibitem{walker} Walker, M. A. 1999, {\it Mon. Not. R. Astron. Soc.}, {\bf 308}, 551
\bibitem{wil89} Willmer, C.N.A., Focardi, P., da Costa, L.N., \& Pellegrini, P.S. 1989, {\it
\AJ}, {\bf 98}, 1531

\end{thebibliography}
\end{document}